\documentclass[prl,nofootinbib,floats,aps,superscriptaddress,twocolumn]{revtex4}
 \pdfoutput=1

\usepackage{amsmath,amssymb,amsfonts}
\usepackage{graphicx}
\usepackage{color}
\usepackage{amsmath}
\usepackage{amsfonts}
\usepackage{amssymb}
\usepackage{float}
\usepackage{cancel}
\usepackage{hyperref}

\newcommand{\beq}{\begin{eqnarray}}
\newcommand{\eeq}{\end{eqnarray}}

\newcommand{\centeron}[2]{{\setbox0=\hbox{#1}\setbox1=\hbox{#2}\ifdim
\wd1>\wd0\kern.5\wd1\kern-.5\wd0\fi \copy0
\kern-.5\wd0\kern-.5\wd1\copy1\ifdim\wd0>\wd1
                                  \kern.5\wd0\kern-.5\wd1\fi}}
\newcommand{\ltap}{\>\centeron{\raise.35ex\hbox{$<$}}
                          {\lower.65ex\hbox{$\sim$}}\>}
\newcommand{\gtap}{\>\centeron{\raise.35ex\hbox{$>$}}
                          {\lower.65ex\hbox{$\sim$}}\>}

\newcommand\ZZ{\hbox{\zfont Z\kern-.4emZ}}
\font\zfont = cmss10 

\newcommand{\cref}[1]{Chapter \ref{c.#1}}

\def\beq{\begin{equation}}
\def\eeq{\end{equation}}
\newcommand{\ba}{\begin{array}}
\newcommand{\ea}{\end{array}}
\newcommand{\bea}{\begin{eqnarray}}
\newcommand{\eea}{\end{eqnarray} }
\newcommand{\bal}{\begin{align}}
\newcommand{\eal}{\end{align}}
\def\bi{\begin{itemize}}
\def\ei{\end{itemize}}
\def\ben{\begin{enumerate}}
\def\een{\end{enumerate}}
\def\beq{\begin{equation}}
\def\eeq{\end{equation}}
\def\bc{\begin{center}}
\def\ec{\end{center}}
\def\bt{\begin{table}}
\def\et{\end{table}}
\def\btb{\begin{tabular}}
\def\etb{\end{tabular}}

\def\co{{\mathcal O}}

\def\gev{\, {\rm GeV}}

\def\mass2{mass${}^2$}

\begin{document}
\pagestyle{plain}  

\title{\boldmath Discovering Higgs Decays to Lepton Jets at Hadron Colliders}

\author{Adam Falkowski}
\affiliation{NHETC and Department of Physics and Astronomy, Rutgers University, Piscataway, NJ 08854}

\author{ Joshua T.~Ruderman}
\affiliation{Department of Physics, Princeton University, 
Princeton, 
NJ 08544}

\author{Tomer Volansky}
\affiliation{School of Natural Sciences, Institute for Advanced Study, 
Princeton, NJ 08540}
\author{Jure Zupan}
\affiliation{Faculty of Mathematics and Physics, 
University of Ljubljana Jadranska 19, 1000 Ljubljana, Slovenia}
\affiliation{Josef Stefan Institute, Jamova 39, 1000 Ljubljana, Slovenia}
\affiliation{SISSA, Via Bonomea 265, I 34136 Trieste, Italy}

\begin{abstract}
  \vskip 3pt \noindent
The Higgs boson may decay predominantly into a hidden sector, producing lepton jets instead of the standard  Higgs signatures.
We propose a search strategy for such a signal at hadron colliders. 
A promising channel is the associated production of the Higgs with a $Z$ or $W$, where the dominant background is $Z$ or $W$ plus QCD jets. 
The lepton jets can be discriminated from QCD jets by cutting on the electromagnetic fraction and charge ratio.  The former is the fraction of jet energy deposited in the electromagnetic calorimeter and the latter is the ratio of energy carried by charged particles to the electromagnetic energy. 
We use a Monte Carlo description of detector
response to estimate QCD rejection efficiencies of order~10$^{-3}$
per jet.  
The expected Higgs mass reach is about 155 GeV at the
Tevatron with 10~fb$^{-1}$ of data and about 135~GeV at the 7~TeV LHC with 1 fb$^{-1}$.
\end{abstract}

\maketitle


{\bf Introduction}. 
The Higgs boson is currently being searched for at the Tevatron and
LHC, and its discovery may well complete the experimental verification of the Standard Model (SM)\@.
Alternatively, the Higgs couplings and branching fractions may differ from the SM predictions. 
In fact, the Higgs couplings to the light SM fermions are predicted to
be very small (e.g. the Yukawa coupling to the bottom quark is $y_b\sim 0.02$). 
The presence of
new light particles can thus drastically change the Higgs decay pattern.    
For this reason, Higgs decays present a promising opportunity for the discovery of new physics.
In Ref.~\cite{Falkowski:2010cm}, we discussed a scenario where the Higgs
boson decays dominantly into {\em  two or more lepton jets plus missing energy}. 
The purpose of this paper is to propose a concrete search strategy for
this Higgs channel at hadron colliders. 

A lepton jet (LJ) is a cluster of highly collimated charged particles: electrons, and possibly muons and pions
\cite{ArkaniHamed:2008qp, Baumgart:2009tn}. LJs can arise, if
there exists a light hidden sector composed of unstable particles with masses in the MeV to GeV range.
A well-motivated class of such models contains a massive vector
particle (a hidden photon) that has a small kinetic mixing with the SM photon~\cite{Holdom:1985ag}.
Due to this mixing, the hidden photon can decay to lighter particles with electric charge.
For example, a $100$ MeV hidden photon decays exclusively to
electrons, whereas a $1$ GeV one decays to electrons, muons
and pions. 
At the Tevatron and LHC, hidden photons and other light hidden
particles are produced with large boosts, causing their visible decay products to form jet-like structures.
This feature makes LJs similar to ordinary QCD jets and the
challenge is to develop experimental techniques that efficiently
isolate the new physics signal from the hadronic background.

As of today, Higgs decays to LJs have not been targeted by any experimental analysis, and the efficiency of existing searches for this sort of signal is low.
The notable exception is the latest LJ search at D0~\cite{newd0}, which constrains the parameter space of
models in Ref.~\cite{Falkowski:2010cm}.  
The D0 search looks for $\Delta R \lesssim 0.2$ clusters,
containing an electron or muon of $p_T > 10$~GeV and at least one companion track
of $p_T > 4$~GeV\@.  
These clusters are required to be isolated in an annulus, $0.2 < \Delta R < 0.4$. 
LJs, however, can be wider than  $\Delta R \simeq 0.2$  and/or
can contain a large multiplicity of leptons with $p_T < 10$~GeV\@.  While the D0 search
is sensitive to narrow LJs with low multiplicities, it would have missed LJs that are wide or more populated,
as can be generic with a non-minimal or strongly coupled hidden sector.
A Higgs boson decaying to such LJs could have escaped all existing searches even if it is very light, $m_h \simeq 100$~GeV~\cite{Falkowski:2010cm}.  
 
In this note we concentrate on Higgs production in association with
 a $W$ or $Z$  and show that the Tevatron or early LHC is sensitive to
Higgs decays to LJs for Higgs masses $\lesssim 155$ GeV\@.  Moreover, we
demonstrate that despite missing energy in the Higgs
decays, it is possible to reconstruct the Higgs mass.  The proposed
search utilizes Higgs-specific kinematic cuts and additional cuts designed to identify LJs with the use of electromagnetic fraction (EMF) and charge ratio (CR)\@. 
EMF is defined as the ratio of jet energy deposited in the
electromagnetic calorimeter (ECAL) to the total jet energy. 
CR is defined as the ratio of the sum of the charged track $p_T$ in the jet to
the transverse energy deposited in the ECAL\@.
 We focus on the scenario where the LJs consist of electrons only (this happens when the hidden photon mass is below
the $2 m_\mu$ threshold). 
In this case the signal has EMF and CR $\simeq 1$, while QCD jets
with EMF near one typically have CR different from 1. 
As we show, combining EMF and CR discriminates lepton jets from
QCD jets, with a background efficiency on the order of  a
$\mathrm{few}\times10^{-3}$ per jet.


{\bf Models}. 
The LJ structure is very sensitive to the details
of the hidden sector.
The signal we study is partially motivated by the weakly coupled models of Ref.~\cite{Falkowski:2010cm}: the MSSM supplemented by a hidden $U(1)_d$ sector consisting of the
hidden photon $\gamma_d$, 2 hidden Higgs scalars and their superpartners. 
The SM Higgs boson decays into the hidden sector
particles, which cascade down, increasing the final state multiplicity. 
At the end of the cascade, the hidden photons decay to electrons
while the lightest hidden fermions carry off missing energy. 
As a result, the Higgs boson decays into 2 or more LJs plus missing
energy.  
Alternatively, LJs can arise from a more complicated hidden sector
(e.g.~with a non-abelian gauge group) or from a strongly coupled
hidden sector  which could result in  even larger final state
multiplicities or wider jet shapes due to showering.

To be able to explore a wide range of LJ collider signatures we use an {\em N-step  cascade} effective model. 
The hidden sector includes the hidden photon $\gamma_d$ mixing with the 
SM photon, a stable scalar $n$ mimicking the lightest hidden
fermion described above, and a set of $N-1$ hidden scalars $h_{d,i}$, that populate the
cascade in the hidden sector.  
The Higgs boson first decays to a pair of hidden scalars $h_{d,1}$,
which then decay to another pair of scalars $h_{d,2}$, and so forth.   
Finally, $h_{d,N-1}$ decays to either a pair of $\gamma_d$ or $n$ and subsequently, the hidden photons decay to pairs of electrons, while $n$  
counts as missing energy.  

The tunable parameters of the effective model include the number of
cascade steps (controlling the electron multiplicity and $p_T$), the
hidden particle masses (controlling the number and width of LJs) and the branching fraction of $h_{d,N-1}$ into $n$ (controlling the amount of missing energy). The effective model is thus flexible enough to simulate the multitude of LJ signatures
available in the parameter space of  \cite{Falkowski:2010cm} and in more 
general hidden sectors.   

In this paper, we present our results assuming a particular 3-step benchmark model. 
The masses of the two unstable scalars are chosen to be 10 and 4~GeV,
while the hidden photon and stable scalar have masses of 100 and 90~MeV, respectively.
The branching fraction of $h_{d,2}$ to $n$ is 20\%.
This benchmark typically produces wide LJs with  $\Delta R  \sim
0.3-0.4$. 
Due to this feature, our benchmark is consistent with the D0 LJ search of
Ref.~\cite{newd0}  for the Higgs mass as low $m_h\sim100$~GeV\@.  
We note that the D0 search has an even lower efficiency for models with longer cascades (more steps), 
such that the leptons are softer than the search's $p_T$ requirement of 10 GeV\@.

\begin{figure*}[t]
\begin{tabular}{cc}
\includegraphics[scale=0.4]{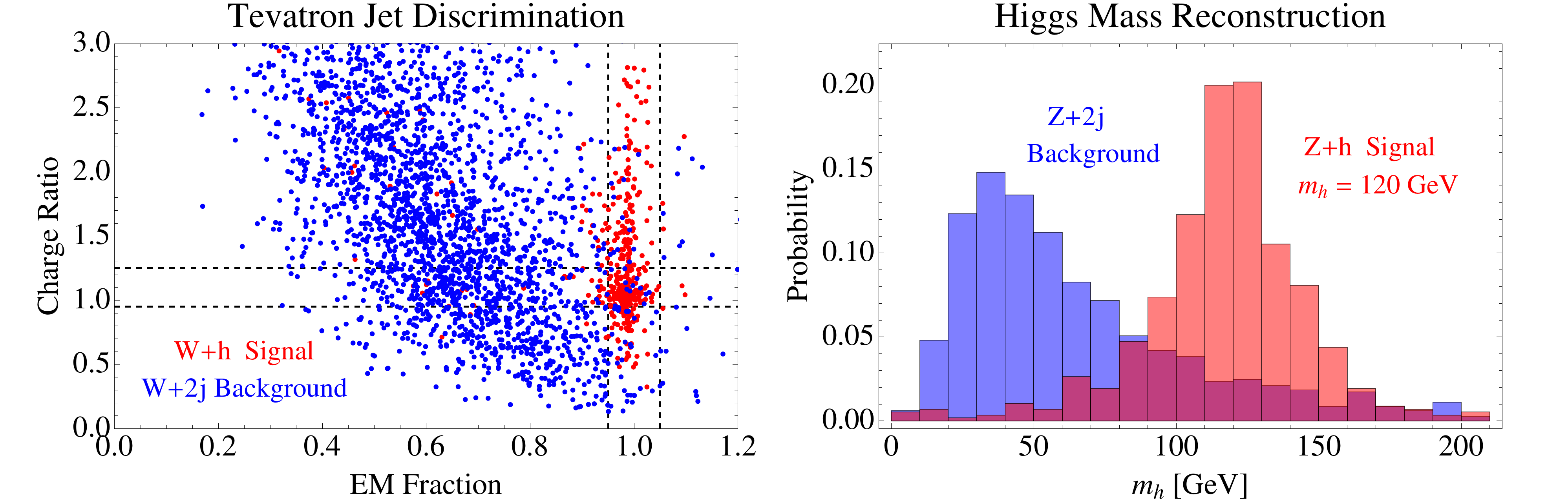}\end{tabular}
\caption{{\it Left:} scatter in electromagnetic fraction (EMF) and charge ratio (CR) for lepton jets (red) and background QCD jets (blue) in the $W$+$h$ channel  at the Tevatron ($m_h = 120$ GeV).  These events have passed the kinematic cuts of Eq.~\ref{eq:k1} and \ref{eq:k2} and the jets have at least 4 tracks.  EMF is the fraction of jet energy deposited in the ECAL and CR is the ratio of the sum of track $p_T$ to the transverse energy deposited in the ECAL.  The signal is clustered at EMF, CR~$\simeq 1$, while these variables are anti-correlated for the QCD background.  The cuts used in the analysis are denoted by dashed lines. {\it Right:} reconstruction of Higgs mass in the $h$+$Z$ channel at the Tevatron for $m_h=120$~GeV, obtained using the approximation that the MET is collinear with the observed lepton jets.  The signal (red) is clearly separated from the $Z+$jets background (blue).
\label{fig:emfpt}
 }
\end{figure*}

{\bf Electron jets vs. QCD jets}.
To discover Higgs decays to LJs we need to tell LJs apart from ordinary QCD jets initiated by
quarks and gluons. 
This is not completely straightforward as closely-spaced leptons  do
not satisfy the usual isolation criteria and will not be reconstructed
as leptons by the experiments.
In Ref.~\cite{Falkowski:2010cm}, we discussed a number of properties 
of LJs that may distinguish them from average QCD jets, e.g.~EMF, jet shapes, and the pair invariant masses of nearby tracks.    
As we show below,  the combination of EMF and CR is a particularly 
powerful discriminating tool that may open the way to a Higgs discovery.   This approach is  orthogonal to the one taken in Ref~\cite{newd0} and captures a different part of the LJ parameter space.

For the signal jets, the electrons typically leave all of their energy in 
the ECAL, so that EMF $\simeq1$.
This gets corrected by occasional leakage of
electromagnetic showers into the HCAL, HCAL noise, or lepton jets
overlapping
with ordinary jets. Nonetheless, most of the signal has EMF
$> 0.95$ (see Fig.~\ref{fig:emfpt}). 

For the background, the picture is more complicated.
By the time a QCD jet reaches the detector, it mainly consists of
charged pions and photons from $\pi^0$ decay.
Most $\pi^\pm$ deposit a sizable fraction of their energy in the HCAL,
while photons deposit almost all their energy in the ECAL\@. 
The precise jet composition, and consequently EMF, fluctuates highly event-by-event.
The distribution is further broadened by fluctuations of the
electromagnetic and hadronic cascades, and by energy
smearing in the detector (the latter also leads to a fraction of jets
having $\textrm{EMF}>1$).  The end result is that the EMF distribution of QCD jets 
peaks around $0.5-0.8$, depending on the detector. A few
percent of jets have EMF $\simeq1$.  
Thus the EMF alone provides only limited discriminatory power. 

The high EMF tail of QCD is due to jets with a high photon
content. These jets leave few tracks and are therefore expected to have small CR\@.  In contrast, LJs composed of electrons have CR $\simeq1$.
The QCD jets and the electron jets are thus
well separated in the EMF-CR plane, as shown in Fig.~\ref{fig:emfpt}.

\begin{figure*}[t]
\begin{tabular}{cc}
\includegraphics[scale=0.41 ]{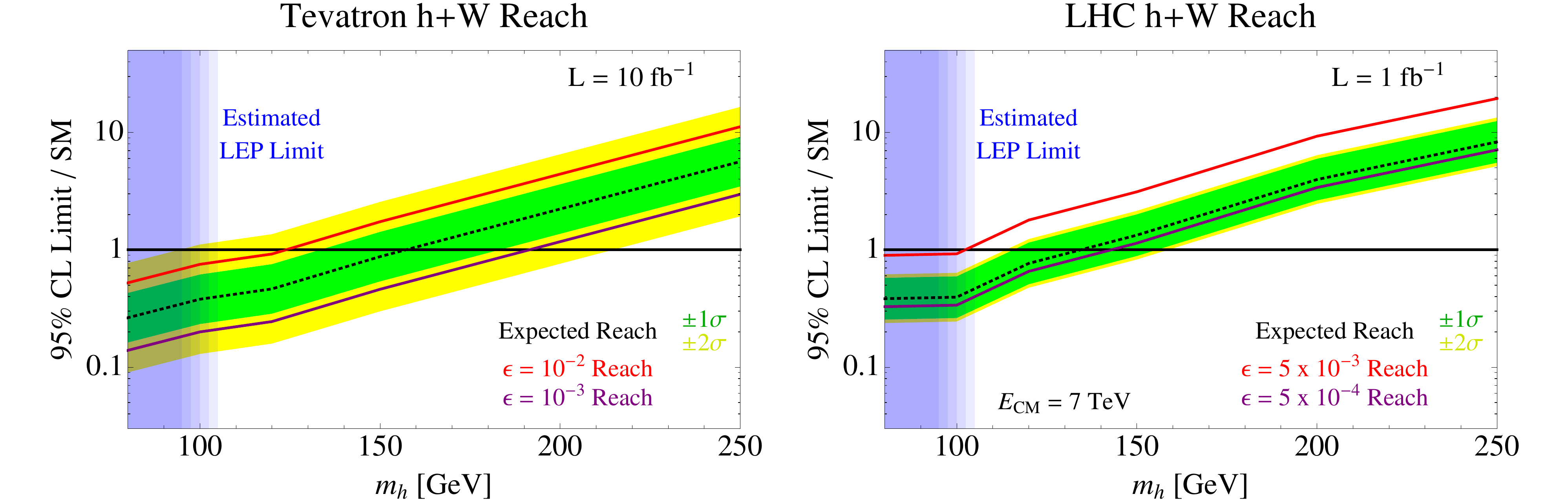}
\end{tabular}
\caption{\label{fig:reach}
Higgs mass reach at the Tevatron ({\it left}) and the early LHC ({\it right}) with luminosities of 10 and 1 fb$^{-1}$, respectively.  The limits are for the $h+W$ channel and are normalized to the SM Higgs production cross-section, assuming a 100\% branching ratio into lepton jets.  The expected 95\% CL exclusion limit (black, dashed) assumes the EMF and CR rejection efficiencies, per QCD jet, extracted from our simulation and shown in Table~\ref{table:finalsignificance}: $\epsilon = 3.7\times10^{-3}$ at the Tevatron and $\epsilon=1.2\times10^{-3}$ at the LHC\@.  The green and yellow bands show the 1$\sigma$ and 2$\sigma$ deviations due to statistical fluctuations of the background.  For comparison, the limits derived from more optimistic (lower) and more pessimistic (higher) values of $\epsilon$ are shown in purple and red, respectively.  Although this signal has not been searched for at LEP, we estimated that the limit is $m_h \simeq 100$~GeV in Ref.~\cite{Falkowski:2010cm}, and this regime is shaded blue.
}
\end{figure*}

{\bf Analysis and Results.} At hadron colliders, the dominant Higgs production mechanism is via gluon fusion, but the
overwhelming dijet background makes this channel very challenging.
Instead, we turn to  
Higgs production in
association with electroweak gauge bosons. 
We search for a leptonically decaying $W$ or $Z$ accompanied by 2 LJs.  The main background is $W/Z$+jets that mimic LJs.

We generated event samples for the D0 detector at the Tevatron and the ATLAS
detector at the LHC with 7 TeV center-of-mass energy. 
Signal and background are generated at the parton level using 
\texttt{MadGraphv4}~\cite{Maltoni:2002qb} and \texttt{BRIDGE}
\cite{Meade:2007js}, and then showered and hadronized in 
\texttt{Pythia 6.4.21}~\cite{Sjostrand:2006za}, including multiple
interactions and pileup.  
The cross-sections are normalized to NLO using MCFM
\cite{Campbell:2002tg}. 
 For detector simulation we use \texttt{PGS4} \cite{PGS} and a private code described below.  
 We first employ kinematic cuts that target the $Z/W$+$h$ signal. 
For the search in the $Z$+$h$ channel we require two opposite sign
same flavor isolated
leptons ($l = e,\mu$) and exactly 2 jets satisfying:  
\beq 
\label{eq:k1}
p_T(j) > 15 \gev, 
\quad 
\Delta R_{j_1,j_2}   > 0.7,
\eeq 
\vspace{-.8cm}
\beq 
\label{eq:k2}
p_T(l) > 10 \gev ,
\quad
|m(l^+l^-) - m_Z| < 10 \gev .
\eeq 
The rapidity cuts are $|\eta| < 2.5$ for D0 (but removing the
$1.1<|\eta|<1.5$ region were ECAL coverage is worse and the
measurement of EMF and CR may be degraded), and 
$|\eta|<2$ for ATLAS for all jets and leptons. 
For the $W$+$h$ channel we use the same cuts on jets, but require one
lepton and missing $p_T$ satisfying, 
\beq 
\label{eq:k3}
p_T(l) > 20 \gev, 
\quad 
p_{T,{\rm miss}} > 20 \gev,
\eeq   
and veto on additional isolated leptons harder than 10 GeV\@.  
The above cuts have efficiency of $\co(10-20\%)$ for the signal,
 see Table \ref{table:finalsignificance}.  

The kinematic cuts are insufficient to overcome the background. We therefore also employ EMF and CR cuts that are targeted at LJs.  We stress that these cuts are not directly related to LJs arising from Higgs decays and would be suitable in any LJ search at hadron colliders.

The \texttt{PGS4}  implementation of calorimeter depositions is too simplistic for
our purpose as it does not take into account realistic EM and hadronic
cascades which are essential for EMF predictions.  We therefore
implement a fast calorimeter simulation for both D0 and ATLAS using a
parametrization of EM showers in sampling
calorimeters~\cite{Grindhammer:1993kw} and the Bock parametrization of
hadronic cascades tuned to D0~\cite{Norman:1993pr} and
ATLAS~\cite{Kulchitsky:2000gg}.  We allow fluctuations of all
parameters and take into account detection efficiency of hadronic and
EM energy (the non-compensation parameter h/e).  Moreover, we simulate
EM energy loss of heavy particles using the Landau-Vavilov
distribution and detector smearing effects tuned to the detectors.
For further details and references, see~\cite{pdg}.    Finally we tune
our simulation, in particular h/e, to D0 and ATLAS EMF data in dijets, obtaining accurate fits.

In order to ensure that our results are not significantly modified by photon
conversions in the tracker, which we do not simulate, we require at least 4 tracks per jet.     
Next we use the code, described above, to estimate the EMF of the signal
and background jets that pass the track cut and  the kinematic cuts (\ref{eq:k1})- (\ref{eq:k3}).   
We estimate the CR of the jets using track $p_T$ from \texttt{PGS4}  divided by jet ECAL deposits obtained
from our code. 
Sample results for $W$+$h$ at the Tevatron are plotted in Fig.~\ref{fig:emfpt}.   
The electron jets are concentrated near EMF, CR $\simeq 1$,
 while the QCD jets display clear anti-correlation of the two
 variables: most of the QCD jets with EMF of order unity have CR different from 1. 
Due to the difference in detector performances, we tune the EMF cut
differently for D0 and ATLAS\@. 
In particular, we find that a tighter EMF cut is required for ATLAS; for D0 we take
$0.95<\mathrm{EMF}<1.05$, while for ATLAS, $0.99<\mathrm{EMF}<1$.  
The CR cut is kept the same for both detectors, but different for the $W$+$h$
and $Z$+$h$ channels. The latter has smaller cross-section and requires looser cuts
to retain enough statistics.   
We take $0.9<\! \mathrm{CR}\!<1.9$ for $Z$+$h$ and $0.95<\! \mathrm{CR} \! <1.25$ for $W$+$h$.

The efficiencies of our kinematic and LJ cuts are summarized in
Table~\ref{table:finalsignificance} for a Higgs, of mass of $120$ GeV,
decaying into LJs modeled by the 3-step cascade described above.   
In Fig.~\ref{fig:reach} we show the Higgs mass reach plot for $10 \textrm{ fb}^{-1}$ of Tevatron data and $1 \textrm{ fb}^{-1}$ of LHC data using the W+h channel.
As can be seen, a $\sim 155$ GeV (perhaps as high as $190$ GeV) Higgs is accessible at the Tevatron, and $\sim135$ GeV Higgs can be probed at the early LHC\@.
The reach is much smaller in the $Z$+$h$ channel due to the smaller cross-section: $\sim110$~GeV at the Tevatron and $\sim80$~GeV at the early LHC\@.
With more LHC data, the reach will improve significantly for both channels.

\begin{table*}
\begin{center}
\begin{tabular}{c|c|cc|cc}
\hline \hline
\multicolumn{2}{c|}{} &\multicolumn{2}{c|}{$W+h$} & \multicolumn{2}{c}{$Z+h$} Ê\\
\cline{3-6}
\multicolumn{2}{c|}{$m_h=120$ GeV} & Ê~~~Signal(Eff.)~~~ & ~~Bckg.~~~& ~~Signal(Eff.)~~ & ~~Bckg.~~~\\
\hline \hline
Tevatron &Kinematic &87 (18\%) &$4.4\times 10^5$ &10.6 (18\%) & $2.8\times 10^4$\\ 
(10 ${\rm fb}^{-1}$)&EMF+CR &14.4 (3\%)& 5.9& 3.5 (6\%)&1.4\\ 
\hline \hline
LHC &Kinematic & Ê35 (17\%) & $4.9 \times 10^5$ &5.2 (25\%) & $3.6\times 10^4$\\  
(1 ${\rm fb}^{-1}$)&EMF+CR &4.9 (2\%)&0.7 &1.5 (7\%) & 0.7\\ 
\hline \hline
\end{tabular}
\end{center}
\caption{The number of signal and background events for the $W$+$h$
  and $Z$+$h$ channels, with $m_h = 120$~GeV, at the Tevatron and LHC\@.
 Event counts are shown after the cuts of Eqs.~\eqref{eq:k1} - \eqref{eq:k3} and requiring at least 4 tracks per jet (Kinematic), and also after including the cuts on
  electromagnetic fraction and charge ratio (EMF+CR). 
\label{table:finalsignificance}
}
\end{table*}

{\bf Higgs mass}.
Finally, we comment that the Higgs mass can be reconstructed from the LJs in the $Z$+$h$ channel.
Although there is missing energy in the final state carried by the $n$'s, we can assume that it is collinear with the two LJs (much like the $h\to \tau\tau$ channel in the SM \cite{Ellis:1987xu}).
This gives 2 unknowns (the magnitudes of the two missing 4-vectors which are taken to be massless), that are fixed by transverse momentum conservation.
The result of applying this procedure is shown in Fig.~\ref{fig:emfpt} for our benchmark model, and the Higgs mass peak is clearly
visible. 
The limiting factor is the small cross-section in the leptonic
$Z$+$h$ channel, which may render the mass reconstruction feasible only for light Higgs mass ($\lesssim 120$ GeV) or with more data.

{\bf Acknowledgments}. 
This work profited greatly from conversations with A.~Atramentov,
 A.~Haas, E.~Halkiades, M.~Pierini, and Y.~Gershtein  and  C.~Tully whom we also thank for guidance with the calorimeter simulation. 
AF was supported in part by DOE grant  DE-FG02-96ER40949, JTR by an NSF graduate fellowship, TV by DOE
 grant DE-FG02-90ER40542,  and JZ by the  EU Marie Curie IEF Grant. no. PIEF-GA-2009-252847 and by the 
Slovenian Research Agency. We thank the Aspen Center for Physics for their hospitality 
in the summer of 2010 where some of this work was completed. 


 \end{document}